\documentclass[iop]{emulateapj}	
\usepackage{natbib,amsmath,hyperref,multirow}
\usepackage[normalem]{ulem}

\newcommand{\Msun}{\mbox{$M_{\odot}$}}
\newcommand{\Mdot}{\mbox{\.M}}

\newcommand{\Mjup}{\mbox{$M_{Jup}$}}

\def\ref{\hangindent=15pt     
        \hangafter=1}

\def\lesssim{\mathrel{\hbox{\rlap{\hbox{%
  \lower4pt\hbox{$\sim$}}}\hbox{$<$}}}}
\def\gtrsim{\mathrel{\hbox{\rlap{\hbox{%
  \lower4pt\hbox{$\sim$}}}\hbox{$>$}}}}

\slugcomment{Accepted, ApJ}

\shorttitle{Millimeter Survey of Upper Sco}
\shortauthors{Mathews et al.}

\begin{document}

\title{The Late Stages of Protoplanetary Disk Evolution:\\ A Millimeter Survey of Upper Scorpius \thanks{Based on observations carried out with the IRAM 30m Telescope as part of programs 066-08, 149-08, and 076-09. IRAM is supported by INSU/CNRS (France), MPG (Germany) and IGN (Spain).}}

\author{Geoffrey S. Mathews\altaffilmark{1}, Jonathan P. Williams\altaffilmark{1}, Francois M\'enard\altaffilmark{2}, Neil Phillips\altaffilmark{3}, Gaspard Duch\^ene\altaffilmark{4,2}, and Christophe Pinte\altaffilmark{2}} 
\affil{Institute for Astronomy, University of Hawaii, Honolulu, HI 96826}
\email{gmathews@ifa.hawaii.edu}

\altaffiltext{1}{Institute for Astronomy, University of Hawaii, 2680 Woodlawn Dr., Honolulu HI 96826}
\altaffiltext{2}{CRNS-INSU / UJF-Grenoble 1, Institut de Plan\'etologie et d'Astrophysique de Grenoble (IPAG) UMR 5274, Grenoble, F-38041, France}
\altaffiltext{3}{Institute for Astronomy, University of Edinburgh, Royal Observatory, Blackford Hill, Edinburgh, EH9 3HJ, UK}
\altaffiltext{4}{Astronomy Department, University of California, Berkeley CA 94720-3411 USA}

\begin{abstract}
We present deep 1.2 millimeter photometry of 37 stars in the young (5 Myr) Upper Scorpius OB association, sensitive to $\sim4\times10^{-3}$ \Mjup ~of cool millimeter dust.  Disks around four low- and solar-mass stars are detected, as well as one debris disk around an intermediate mass star, with dust masses ranging from $3.6\times10^{-3}$ -- $1.0\times10^{-1}$\Mjup.  The source with the most massive disk exhibits a transition-disk spectral energy distribution.  Combining our results with previous studies, we find the millimeter-detection fraction of Class II sources has significantly decreased from younger ages, and comparison with near-infrared and H$\alpha$ measurements indicates the present disks have undergone significant evolution in composition or structure at all radii.  The disks of Upper Scorpius represent the tail-end of the depletion of primordial disks; while a few near-solar mass stars may still sustain giant planet formation, this process has finished around higher mass stars.

\end{abstract}


\section{Introduction}
\label{sec:intro}

Our understanding of circumstellar disks, the birthplace of planets, has been driven in the last decade largely by the Spitzer space telescope, which has allowed for the identification of young stars with warm, optically thick inner dust disks.  Spitzer has observed all known nearby sites of recent star formation (e.g. c2d, FEPS, Gould's Belt Survey), revealing hundreds of new disk bearing systems with ages from 1 -- 100 Myr.  These disks have shown a remarkable variety of spectral energy distributions (SED), indicating the presence of large-excess primordial disks, weaker excess ``anemic'' disks, transition disks showing an emission gap at short wavelengths, and debris disks showing weak emission characteristic of small amounts of second generation dust \citep[for an overview, see][]{Williams:2011}.  These many surveys have revealed a picture of circumstellar disk evolution in which inner disk dust dissipates on timescales of a few Myr \citep[e.g.][]{2005astro.ph.11083H}, with inner disks around higher mass stars evolving more quickly \citep[e.g.][]{2006ApJ...651L..49C}.  

Optical and near-infrared spectroscopic surveys for accretion indicators (e.g. H$\alpha$ line equivalent width) have revealed a similar variety in accretion rates among young stars \citep[e.g.][]{Dahm:2008}.  The presence of accretion indicates the existence of gas in the inner disk.  

On the other hand, probing the outer disk, the main reservoir of dust, requires sensitive (sub-)millimeter photometry.  At these wavelengths, the dust is optically thin, allowing for the direct measurement of dust mass.  Past large mm-surveys in young ($<$10 Myr) regions have reached dust mass sensitivities of a few $10^{-2}$ \Mjup ~for standard assumptions \citep[e.g.][]{,2005ApJ...631.1134A,2009A&A...497..409R}.  In this work, we present results of a sensitive search for millimeter continuum in a sample of nearby stars in the Upper Scorpius region which have signs of significant evolution in the inner disk dust and gas.  

At a distance of 145 pc \citep{de-Zeeuw:1999}, \object{Upper Scorpius} (Upper Sco) is one of the closest sites of recent star formation.  At an age of 5 Myr \citep{Preibisch:1999}, it has largely exhausted its interstellar gas and dust, reducing contamination from field emission.  Spectroscopic studies indicate that few stars are still accreting \citep[e.g.][]{Walter:1994,2009AJ....137.4024D}; this suggests that most disks have depleted their inner disk gas content.  Spitzer studies using IRAC, IRS, and MIPS \citep{2006ApJ...651L..49C,2009ApJ...705.1646C} directly revealed that the disks present in Upper Sco are more evolved than disks in younger star forming regions.  The majority of Upper Sco members ($\sim$80\%), at all spectral types, exhibit no infrared excess.  SEDs for the infrared detections in Upper Sco indicate anemic or transition disks around K and M stars, and debris disks around stars across the entire sample (B through M spectral types).  Similar results have been found in other 5 Myr associations \citep[e.g. $\lambda$ Ori and NGC 2362,][]{Barrado-y-Navascues:2007,Dahm:2007}.

The variety of disk types around low mass stars in Upper Sco suggest that an age of 5 Myr may be a time at which disk properties rapidly change, providing an ideal laboratory for examining the evolution of disks.  Of all mechanisms affecting disk evolution, perhaps the one that has stirred the greatest interest in recent years has been the possibility of disruption by giant planets.  Disk mass measurements will provide direct constraints on the potential for on-going giant planet formation, in turn constraining planet formation models.  

We have carried out the most sensitive survey to date for millimeter emission from 5 Myr stars, reaching a dust mass sensitivity of $\sim$4$\times10^{-3}$ \Mjup.  In section \ref{sec:observations}, we describe our target sample and our observations.  In section \ref{sec:results}, we discuss the inferred properties of our sources.  In \S \ref{sec:discussion}, we discuss the properties of disks in Upper Sco, and in \S \ref{sec:evolution}, we compare to disks at younger ages and discuss the implications for planet formation.  We summarize our findings in section \ref{sec:summary}.

\section{Sample, observations, and values from the literature}
\label{sec:observations}

We selected our sample from the 218 Upper Sco members observed in the near-infrared (NIR) continuum survey of \cite{2006ApJ...651L..49C}.  They selected survey targets that were identified as group members on the basis of stellar parameters so as to not bias towards disk bearing sources.  Moreover, their targets were chosen so as to uniformly sample the stellar mass range in Upper Sco.  We thus believe their sample to be the best representation of the Upper Sco population.  Our sample focused on sources identified by \citeauthor{2006ApJ...651L..49C} as having a NIR excess at 8 or 16 $\mu m$; we observed 8 of the 9 B and A (henceforth BA) excess sources, 20 out of 24 K and M stars (henceforth KM) with an 8 or 16 $\mu m$ excess, and 3 BA and 6 KM stars with no excess.  

Our targets were observed for continuum emission at 1.2 mm using the MAMBO2 bolometer array \citep{1998SPIE.3357..319K} on the IRAM 30m telescope at Pico Veleta, Spain.   Observations were conducted during the bolometer pools in 2008 (Nov. 13-18, 23) and 2009 (February 24-25, March 8-9, and Nov. 8).  Zenith opacity for our observations was typically $\sim$0.2 -- 0.3.  Observations were carried out to a target 1$\sigma$ sensitivity of 1 mJy, typically 20 minutes on source, in an ON-OFF pattern of 1 minute on target followed by 1 minute on sky, with a throw of 32\arcsec.  Observations during November 2009 were limited to 10 minutes on source, resulting in poorer sensitivity.  In addition, deeper observations of objects showing 2$\sigma$ detections at the target RMS were carried out, resulting in the discovery of millimeter emission from a pair of sources (HIP 76310, ScoPMS 31) at a higher sensitivity than the median of the sample.  Flux calibration was carried out using Mars, Saturn, or Jupiter (depending on availability), and local pointing and secondary flux calibration was carried out using IRAS 16293-2422B.  The data were reduced using the facility reduction software, MOPSIC\footnote{http://www.iram.es/IRAMES/mainWiki/CookbookMopsic}.  

We list the physical properties of the sources in this study in Table \ref{tab:prop}.  We include values for three objects from other studies: \object{$[$PBB2002$]$ USco J160823.2-193001} and \object{$[$PBB2002$]$ USco J160900.7-190852} \citep[observed at 1.3 mm in][]{Cieza:2008}, and \object{$[$PZ99$]$ J161411.0-230536} \citep[observed at 1.2 mm in][]{2009A&A...497..409R}.  These are members of the parent sample from \cite{2006ApJ...651L..49C}, and would have been observed as part of this work if photometry did not already exist in the literature.

\subsection{Spectral type and stellar mass}
\label{ref:stellar_mass}

The spectral types of BA stars are from \cite{Hernandez:2005}, and for KM stars we use spectral types from \cite{1998A&A...333..619P,Preibisch:2002}.  For BA stars, we adopt stellar masses from \cite{Hernandez:2005}, who used the stellar isochrones of \cite{Palla:1993}.  For KM stars, we use masses from \cite{2007ApJ...662..413K}, who used the models of \cite{DAntona:1997} and \cite{Baraffe:1998} for \textbf{solar-mass} and low mass stars, respectively.  Two stars lack a mass estimate in the literature, $[$PBB2002$]$ USco J160823.2-193001 and $[$PBB2002$]$ USco J160900.0-190836; we assigned these stars masses according to their spectral type, as in \cite{2007ApJ...662..413K}.  Comparison of the KM star masses to other estimates in the literature \citep[e.g.][]{Preibisch:1999} suggests a typical stellar mass uncertainty of 10--20\%.  Both spectral type and estimated stellar mass are listed in Table \ref{tab:prop}.  

\subsection{Accretion}
\label{ref:accretion}

The H$\alpha$ equivalent width, W$_{\lambda}$(H$\alpha$), was taken from \cite{Hernandez:2005}, \cite{1998A&A...333..619P,Preibisch:2002}, \cite{2009AJ....137.4024D}, and \cite{Riaz:2006}.  This is then used to determine the accretion state of the stars.  Traditionally, KM stars with W$_{\lambda}$(H$\alpha$) of 10\AA ~or greater were considered to be accreting, and referred to as Classical T-Tauri stars (CTTS).  Those with lower or undetected emission were considered to be non-accreting, and referred to as Weak-line T-Tauri stars (WTTS).  Recent work has refined this classification system \citep{Barrado-y-Navascues:2003,White:2003}, identifying spectral type dependent W$_{\lambda}$(H$\alpha$) that unambiguously indicate accretion.  Here, we adopt the system of \citeauthor{White:2003} --- W$_{\lambda}$(H$\alpha$) in emission of 3, 10, 20, and 40\AA ~indicate accretion for K0--K5, K7--M2.5, M3--M5.5, and M6 and later spectral types, respectively.  While there are a few accreting systems among the KM stars of Upper Sco, none of the BA stars in Upper Sco show H$\alpha$ in emission, suggesting that none of these stars still experience accretion. 

We note that the use of an accretion indicator to demonstrate the presence of gas in the disk is subject to two main problems: accretion and photospheric H$\alpha$ emission can be difficult to distinguish at low equivalent widths, and accretion may vary by large amounts on timescales as short as days \citep[][]{Bouvier:2007a,Nguyen:2009}.  Among the 20 KM stars in our sample with multiple epochs of W$_{\lambda}$(H$\alpha$) measurements, two show values which would produce different classifications as CTTS or WTTS over a decade timescale.  In these two cases ($[$PBB2002$]$ J160545.4-202308 and $[$PBB2002$]$ J160702.1-201938), we classify the sources as CTTS.  We list the W$_{\lambda}$(H$\alpha$) and accretion state of our sources in Table \ref{tab:prop}.

Our observations include 8 of the 10 CTTS in the original sample of \cite{2006ApJ...651L..49C}, and 17 of the 95 WTTS.  There are an additional 22 KM stars with no H$\alpha$ measurement in the literature, of which we include 1.  The millimeter-detected sources we include from the literature are split between accreting and non-accreting systems; 1 is a CTTS, and 2 are WTTS. 

\subsection{Infrared disk classification}
\label{sec:IR_disk}

Table \ref{tab:prop} also shows the disk classification based on SEDs.  SEDs were constructed using 2MASS J, H, and K band photometry \citep{Skrutskie:2006}, 4.5, 8.0, and 16 $\mu$m photometry from \cite{2006ApJ...651L..49C}, and 24 and 70 $\mu$m photometry from \cite{2009ApJ...705.1646C}.  We used extinction estimates from the literature \citep{Hernandez:2005, 1998A&A...333..619P,Preibisch:2002} and the R=3.1 reddening law of \cite{Fitzpatrick:1999} to deredden NIR photometry.  

For the BA stars in Upper Sco, we use the SED classifications of \cite{2009ApJ...705.1646C}.  Among BA stars, they found only debris disks.  \citeauthor{2009ApJ...705.1646C} classified stars in Upper Sco as hosting debris or primordial disks on the basis of the strength of the NIR excess and comparison of NIR colors with those of disks in the younger IC348 region.  

For KM stars, the classification is based upon the slope of the SED at NIR wavelengths, as in \citep{Greene:1994a}.  Assuming $\lambda F_{\lambda} \propto \lambda^n$, sources exhibiting a near infrared power law index greater than 0.3 are considered Class I, $-0.3 < n < 0.3$ are flat-spectrum, $-1.6 < n < -0.3$ are Class II, and $n < -1.6$ are Class III.  The classes roughly trace the expected evolution of inner disk dust.  We use the broadest baseline applicable to our entire sample, calculating the index from K-band and 24 $\mu m$ photometry. 

We classify the K2 star $[$PZ99$]$ J160421.7-213028 as hosting a transition disk, based on its lack of NIR excess at wavelengths less than 16 $\mu m$ and steeply rising excess at longer wavelengths \citep[e.g.][]{Cieza:2007}.  

The Upper Sco sample of \cite{2006ApJ...651L..49C} contains 19 Class II sources, comprising 15\% of the KM stars, and 107 Class III sources.  We observed 16 Class II and 9 Class III sources.  The three additional detections included from the literature are Class II sources.

Due to the variety of disk types among KM stars, we summarize their disk classifications in Table \ref{tab:disk_classification}, where we show the number of each type of object in the parent sample of \cite{2006ApJ...651L..49C} (applying our Class II / III and CTTS / WTTS criteria as described above), the number included in this work, and the number of millimeter detections.

\section{Results and analysis}
\label{sec:results}

We detect 1 out of 11 BA stars, and 4 out of 26 KM stars.  Our median 3$\sigma$ sensitivity is 2.8 mJy.  We report observation times, fluxes, and 3$\sigma$ upper limits in Table \ref{tab:obs}.  We also include fluxes from the literature for $[$PBB2002$]$ USco J160823.2-193001, $[$PBB2002$]$ USco J160900.7-190852, and $[$PZ99$]$ J161411.0-230536. 

Overall, millimeter observations exist for 9 of the 10 CTTS in the original sample, of which 4 are detected.  18 of the 95 WTTS have been observed for millimeter emission, with 3 detections, and the single millimeter-observed M star lacking H$\alpha$ information is not detected. 

Examining disks by NIR classification, 6 Class II sources are detected in millimeter emission.  The additional two millimeter-detections are the debris disk of HIP 76310, and the transition disk of $[$PZ99$]$ J160421.7-213028.

\begin{figure*}
\vspace{2cm}
  \includegraphics[angle=0,width=0.65\columnwidth]{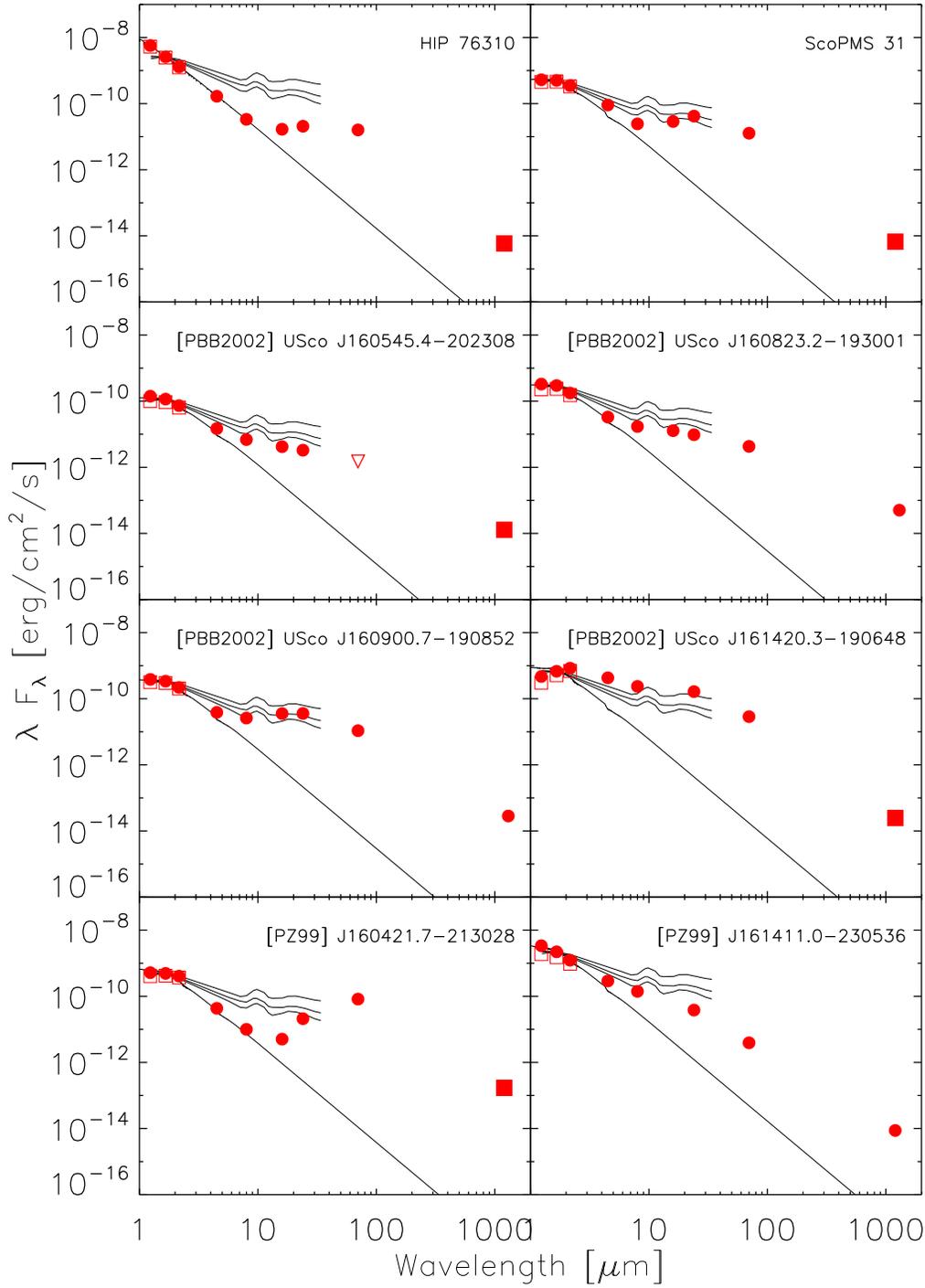}
  \caption{SEDs of the millimeter detected sources discussed here, shown with Kurucz photospheric models \citep{Kurucz:1979} and overlaid with the upper quartile, median, and lower quartile SEDs of Taurus KM stars \citep{Furlan:2006} normalized to the H band flux.  Red circles are dereddened fluxes from the literature \citep[R=3.1,][]{Fitzpatrick:1999}, with open squares indicating original measurements.  Our measurements are indicated as filled red squares.  Inverted triangles indicate upper limits.  1$\sigma$ errors are smaller than symbols.}
  \label{fig:sed1}
\end{figure*}

We show the SEDs of the 8 millimeter detected Upper Sco sources in Figure \ref{fig:sed1}.  We also display the upper quartile, median, and lower quartile SEDs of disk bearing sources in the younger Taurus association \citep{Furlan:2006} for comparison.  Most sources lie in the lower quartile of disk emission in Taurus, implying extensive evolution or settling of the dust.  For the object $[$PBB2002$]$ USco J161420.3-190648, the photospheric model appears to be a poor fit at short wavelengths.  This source is likely being viewed close to edge on.

\subsection{Mass estimates}
\label{sec:mass}

Dust masses were estimated from the 1.2 mm flux density assuming the dust to be optically thin, and are listed in Table \ref{tab:obs}.  Under this assumption, the dust mass ($M_{dust}$) will be directly proportional to the flux (F$_{\nu}$), as in \cite{Hildebrand:1983}:

\begin{equation}
M_{dust} = \frac{d^2F_{\nu}}{\kappa_{\nu}B_{\nu}(T_c)}
\label{eqn:mdust}
\end{equation}

Assuming a dust mass opacity ($\kappa_{\nu}$) of 10 cm$^2$/g at 1000 GHz and opacity power law index $\beta$=1 \citep{Beckwith:1990}, as well as a characteristic dust temperature ($T_c$) of 20 K \citep[the median temperature of Taurus disks; ][]{2005ApJ...631.1134A}, we can then estimate dust masses as $M_{dust} = 1.5\times10^{-3} F_{1.2mm}$(mJy) \Mjup at the 145 pc distance of Upper Sco.  For the two objects observed at 1.3 mm, $[$PBB2002$]$ USco J160823.2-193001 and $[$PBB2002$]$ USco J160900.7-190852, we must adopt a different constant: $M_{dust} = 1.9\times10^{-3} F_{1.3mm}$(mJy) \Mjup.  While there is considerable theoretical uncertainty in these terms, the values are consistent with previous disk studies.  The addition of far-infrared photometry will allow for detailed modeling of the dust structure of the disk and an improved dust mass estimate \citep[e.g.][]{Pinte:2006}, while multi-line spectroscopy is necessary to constrain the gas mass \citep[e.g.][]{Kamp:2011}.  These are the subject of a future paper.

Our median 3$\sigma$ sensitivity of 2.8 mJy corresponds to an estimated dust mass sensitivity of 4.2$\times10^{-3}$ \Mjup.  In order to estimate an upper limit to the average dust mass for non-detected BA and KM stars, we stack our non-detections.  If many sources have mm-emission just below our 3$\sigma$ limits, then we would expect stacking to result in a measurable flux.  However, this procedure results in a non-detection for both the stacked BA and KM observations.  The 3$\sigma$ upper limits of the stacked observations, 0.89 mJy for BA stars, and 0.73 mJy for KM stars, do not differ from what would be expected from the noise in a single observation of the combined observation times.  These upper limits suggest that on average, BA and KM stars have dust masses less than 1.3$\times10^{-3}$ and 1.1$\times10^{-3}$ \Mjup, respectively.

\section{Discussion}
\label{sec:discussion}

\subsection{Comparison of detected and non-detected sources}
\label{sec:diskproperties}

\begin{figure}
  \includegraphics[width=0.95\columnwidth]{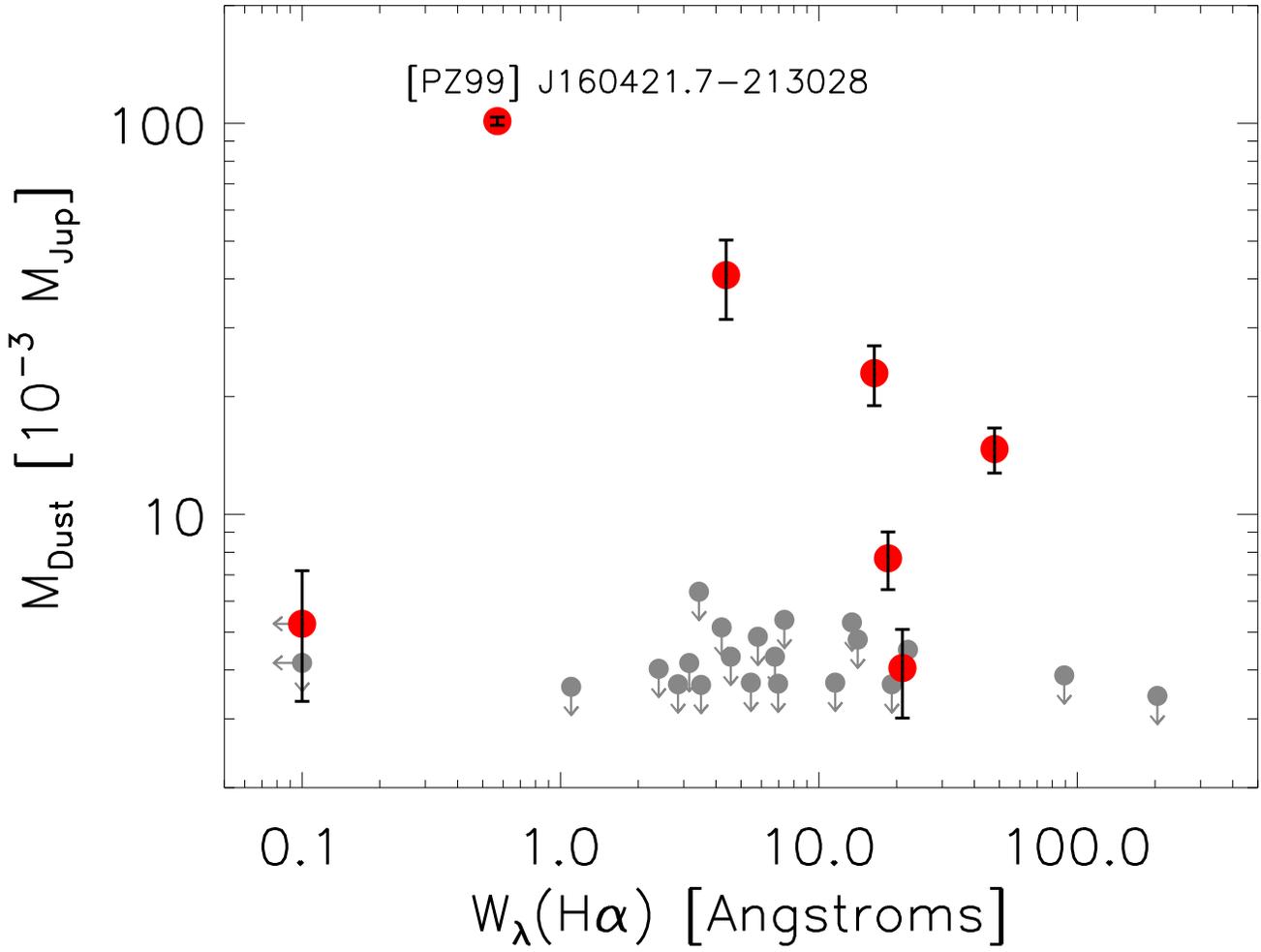}
  \caption{Estimated disk mass versus mean H$\alpha$ equivalent widths for millimeter detected (red) and non-detected KM stars (gray) in Upper Sco.  The displayed errors on $M_{dust}$ only include flux errors; we neglect uncertainties due to adopted opacity and average disk temperature.  There is no overall correlation between disk mass and $W_{\lambda}(H\alpha)$.  }
  \label{fig:mdust_vs_halpha}
\end{figure}

The millimeter-detected KM stars exhibit H$\alpha$ emission similar to the non-detected KM stars.  For each star, we calculate the mean H$\alpha$ equivalent width from values in the literature, and show these in comparison to the dust masses in Figure \ref{fig:mdust_vs_halpha}.  We show sources with H$\alpha$ in absorption as having an upper limit to emission of 0.1 \AA.  We calculate the Kendell's rank statistic, generalized for a censored data set \citep{1986ApJ...306..490I}, for the KM stars and find a Z-value of 0.028, indicating a 98\% chance there is no correlation between the H$\alpha$ equivalent widths of Upper Sco KM stars and their disk dust masses.  Considering just the mm-detected sources, there is ambiguous evidence for an anti-correlation (Z-value of 1.05, 29\% chance of no correlation).

\begin{figure}
  \includegraphics[width=0.95\columnwidth]{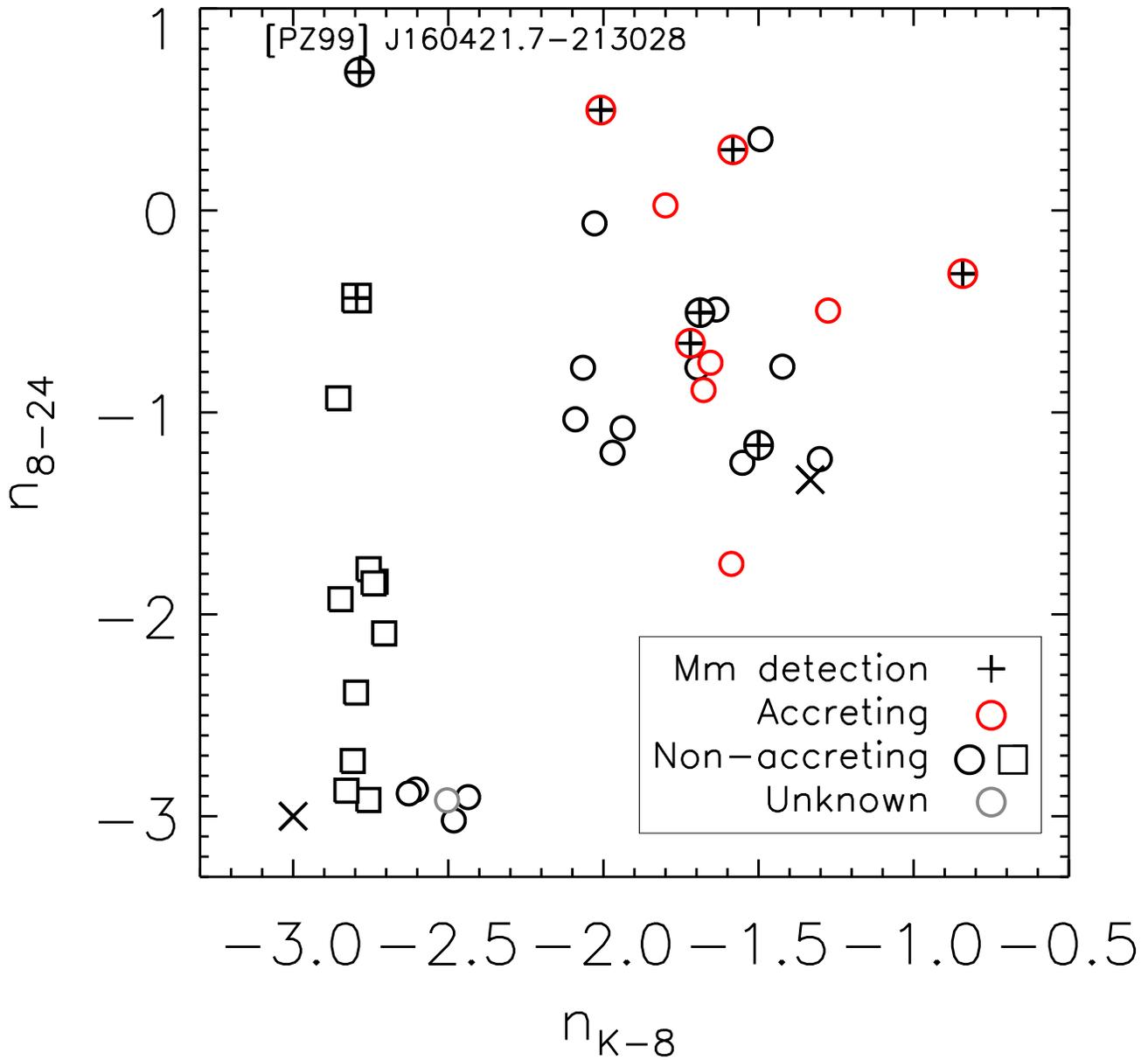}
  \caption{Near-- and mid--infrared SED indices, following the example of \cite{Furlan:2006}.  Squares indicate BA sources, circles indicate KM sources, and plus signs indicate millimeter detected sources.  The symbol color indicates accretion status, and ``$\times$'' marks in the center and bottom left indicate the expected indices for a uniform flat disk and a stellar photosphere, respectively.  The majority of detections have similar SED shapes to other Upper Sco Class II sources. }
  \label{fig:SED-slopes}
\end{figure}

To carry out detailed comparison of the SED shapes, we plot the SED indices from 2.2 to 8 microns and 8 to 24 microns in Fig. \ref{fig:SED-slopes}, defining the SED index $n_{1-2} = \log(\frac{\lambda_2F_{\lambda_2}}{\lambda_1F_{\lambda_1}}) / \log(\frac{\lambda_2}{\lambda_1})$ as in \cite{Furlan:2006}.  The central cluster of points correspond to Class II sources, while points along the left hand side show no excess at 8 $\mu m$, indicative of a void inner region.  However, some of these sources exhibit 24 $\mu m$ excesses  indicative of some amount of cooler dust which may either represent transition or debris disks.  The majority of millimeter-detected sources in Upper Sco exhibit NIR SEDs typical of other Upper Sco Class II sources.

\begin{figure}
  \includegraphics[width=0.95\textwidth]{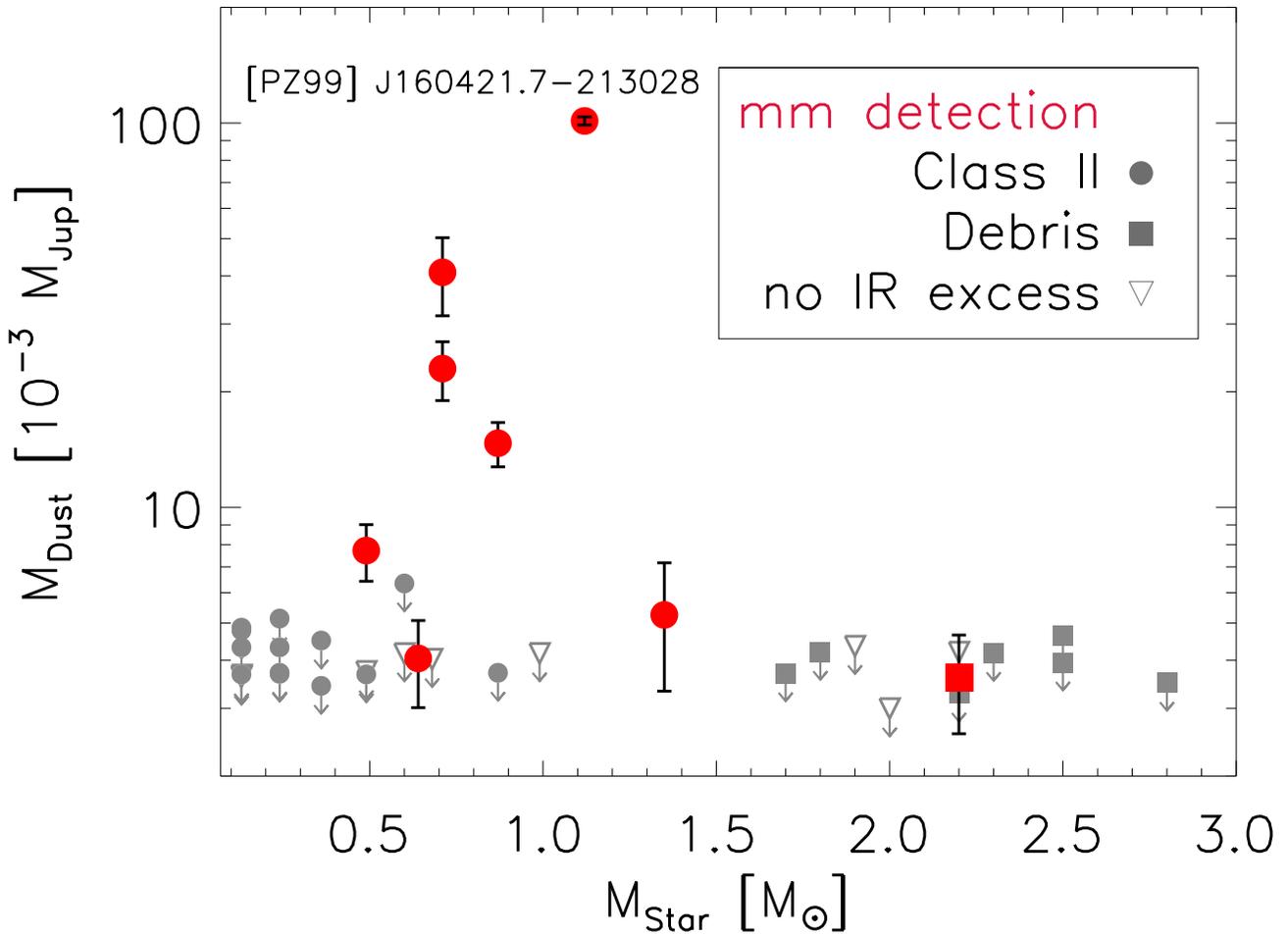}
  \caption{Estimated disk dust mass vs. stellar mass in Upper Sco (5 Myr).  Solid circles indicate Class II or transition disks, solid squares indicate sources with debris disk-like IR excesses, and inverted triangles indicate non-excess sources.  Millimeter detected sources are red symbols, with 1$\sigma$ error bars.}
  \label{fig:mdiskmstar}
\end{figure}

While the NIR and H$\alpha$ properties of the majority of millimeter detected sources do not stand out among Upper Sco source, disk dust mass does appear as if it could be correlated with stellar mass.  In Figure \ref{fig:mdiskmstar}, we show the dust mass as a function of stellar mass.  The highest dust masses appear clustered around stars of 0.7--1.2 \Msun, suggesting a stellar mass dependence in the evolution of the mass of millimeter-sized grains in disks.

We calculate the Kendall's rank correlation statistic and find that the comparison is ambiguous when considering the entire sample.  With a Z-value of 0.87, there is a 39\% chance of rejecting the null-hypothesis that the disk mass and stellar mass are unrelated.  Noting that the BA stars only host debris disks, we examine the stellar mass - dust mass relation for KM stars.  In this case, the relationship is clear; with a Z-value of 3.17, there is only a 0.2\% chance of disk dust mass not being related to the stellar mass.  In Upper Sco, K stars tend to have disks with higher dust mass than do M stars.  This could reflect numerous factors; these higher mass stars may have emerged from their natal clouds with higher mass disks or some mechanism may be active in this mass range to preserve disks.

The observation that millimeter detected sources are not outliers in their NIR and H$\alpha$ emission suggests that the process leading to low millimeter emission is not necessarily tied to the loss of disk gas nor small warm grains.  However, the concentration of high dust masses around near-solar mass stars suggests a stellar mass dependent mechanism in dust evolution.  This expands upon the results of \cite{2006ApJ...651L..49C} and \cite{2009ApJ...705.1646C}, who found evidence that small dust grains deplete more rapidly around stars of greater than solar mass than around stars of near-solar mass or less; our results suggest millimeter-emitting grains are retained longer around near-solar mass stars.

\subsection{Effects of detectability}
\label{sec:detectability}

The NIR and H$\alpha$ classification systems we use above serve as indicators for the presence of dust and gas components of disks.  While in younger groups the two properties are strongly correlated, in Upper Sco  the majority of Class II systems show no evidence for accretion.  The statistics of our observations and data collected from the literature indicate that we are seeing the tail-end of the primordial disk stage.  All disks hosted by BA stars appear to be debris disks.  Only 22 out of 127 (17\%) KM stars show any evidence for remaining primordial disk material (either NIR excess, H$\alpha$ emission, or mm-emission), and only 7 (6\%) have disks detected in millimeter photometry.   
 
These remaining primordial disks appear to be in the last stages of dissipation --- their SEDs lie below the Taurus median, indicating significant inner disk dust evolution or flattening of the disk; they generally have low H$\alpha$ equivalent widths, suggesting depleted gas reservoirs for sustained accretion; and their low mm fluxes indicate that the mass in millimeter-sized and smaller grains has, for the most part, decreased to the edge of detectability with current facilities.

One must keep in mind, however, that each of these measures reaches a different mass sensitivity.  Because of this, one should exercise caution in interpreting the statistics regarding Class II / Class III sources and CTTS / WTTS sources.  

Our millimeter observations reach a dust mass sensitivity of $\sim4\times10^{-3}$\Mjup, while near-IR observations are sensitive to dust masses of $\sim10^{-5}$ \Mjup ~\citep[e.g.][]{Cieza:2007}.  These correspond to disk mass sensitivities of 0.4 \Mjup ~and $10^{-3}$ \Mjup, assuming the ISM gas-to-dust ratio of 100 \citep[e.g.][]{Beckwith:1990}.

To estimate an equivalent disk mass sensitivity for H$\alpha$ observations, we use the empirical relation of accretion rate as a function of time of \cite{Hartmann:1998}.  Modeling the accretion rate ($\Mdot$) as a power law in time ($\Mdot \propto t^{-\eta}$, with stellar age $t$ and $\eta \approx 1.5-2.8$), one can estimate the remaining disk gas mass as $M_{gas} \approx (0.5 - 2)\times\Mdot(t)\times t$.  The W$_{\lambda}$(H$\alpha$) cutoffs for classifying systems as accreting roughly correspond to accretion rates of $10^{-10}$\Msun/yr \citep[e.g.][]{Dahm:2008}.  Thus, for our Upper Sco sources with $t = 5$ Myr, W$_{\lambda}$(H$\alpha$) observations provide an estimated disk mass sensitivity of $\sim(0.3-1)\times10^{-3}$\Msun.  While this estimate spans a large range of values, and neglects many of the details of accretion (e.g. accretion variability, observation geometry), it illustrates that H$\alpha$ observations are likely sensitive to disks of similar mass as those detected by our millimeter photometry.  Both W$_{\lambda}$(H$\alpha$) and millimeter photometry based estimates of the disk mass are at least 2 orders of magnitude less sensitive than near-IR observations.  

Therefore, one should expect the presence of numerous systems exhibiting infrared excesses, but lacking one or both of detectable H$\alpha$ emission and detectable millimeter emission.  This is indeed the case in Upper Sco, with only 4 sources exhibiting all three disk tracers, while 80\% of accreting sources and all millimeter detected sources exhibit an excess at 24 $\mu m$ and shorter wavelengths.

\subsection{Notes on selected sources}
\label{sec:notes}

\label{sec:HIP76310}
\textbf{HIP 76310:}  The sole mm-detected A star in our sample, HIP 76310, has the brightest infrared excess of the BA stars in Upper Sco.  However, its infrared to stellar luminosity ratio \citep[$L_{IR}/L_{*} = 2.1\times10^{-4}$, ][]{2009AJ....137.4024D} falls within the observational definition of a debris disk ($L_{IR}/L_{*} < 10^{-3}$).  Our assumption of 20K for the average dust temperature of primordial disks is based on KM stars in Taurus and does not apply to the case of a debris disk around an A star.  The SED fit by \citeauthor{2009AJ....137.4024D} implies a much higher dust temperature of 122K and thus a correspondingly lower dust mass (a factor of $\sim$8).  Both our initial mass estimate ($3.6\times10^{-3}$\Mjup) and revised dust mass of $\sim5\times10^{-4}$ \Mjup ~are several orders of magnitude higher than the lower limit of $2.6\times10^{-7}$\Mjup ~found by \citeauthor{2009AJ....137.4024D}, highlighting the high NIR optical depth achieved by small quantities of dust, and the importance of (sub-)millimeter observations to directly constrain the dust mass.  This dust mass also fits the evolutionary trend of debris disk dust masses as a function of time as shown in \cite{Williams:2006}.  

\label{sec:J1604} 
\textbf{$[$PZ99$]$ J160421.7-213028:}  The K2 star $[$PZ99$]$ J160421.7-213028 stands out not only for having a uniquely steep infrared SED in Upper Sco, but also for having the largest millimeter emission.  The SED shape for this object is typical of a transition disk, where dust grains have been cleared from the inner disk but are still abundant at larger radii.  The strong millimeter emission implies a large dust mass (0.10 \Mjup).  However, this source is not clearly accreting; the H$\alpha$ equivalent width of -0.57 \AA ~suggests that some process has greatly reduced or cut off accretion to the stellar surface.  By current disk evolution models, the two most likely candidate processes are the onset of photoevaporation \citep{Alexander:2006} or the formation of a planet large enough to reduce stellar accretion below our detection limits \citep{2007MNRAS.375..500A}.  Detailed examination of this disk requires millimeter imaging, which we defer to a later paper.  
 
\label{sec:binaries}

\textbf{Binaries:}  ScoPMS 31 and [PZ99] J161411.0-230536 are detected in millimeter continuum and exhibit primordial type infrared excesses, yet have binary companions at projected separations of 84 and 32 AU, respectively \citep{Kohler:2000,Metchev:2009}.  There is ambiguity between observing a circumstellar or circumbinary disk; both cases allow for the presence of cold grains at large radii.  While the question of the effect of binarity on disk mass and lifetime is worth considering, our numbers here are too small to provide a useful study.

\section{Disk evolution}
\label{sec:evolution}

We can investigate the evolution of observable disk properties from 1 to 5 Myr by comparing the properties of KM sources in Upper Sco with those in Taurus \citep[mean age 1 Myr, ][AW05]{2005ApJ...631.1134A}, and in IC348 \citep[mean age $\sim$2.5 Myr;][LWC11]{Lee:2011}.  These studies are among the largest millimeter surveys at younger ages, for populations that have served as benchmarks in many disk studies.  The Upper Sco and Taurus samples are readily comparable, as our study has similar dust mass limits to AW05 ($4\times10^{-3}$\Mjup ~versus $6\times10^{-3}$\Mjup).  Comparisons with IC348 (dust mass limit of $2\times10^{-2}$\Mjup) will necessarily focus on the high mass end of the dust mass distributions.  We examine trends in disk mass as a function of disk accretion and NIR excess, and the implications for planet formation.

\subsection{Accretion status}
\label{sec:Halpha}

At younger ages accreting systems dominate the population of sources with millimeter continuum detections.  In Taurus, 67 out of 76 (88$\pm$11\%) millimeter detected sources with H$\alpha$ measurements in AW05 are accreting, while in IC348, all 9 mm-detected sources are accreting (LWC11).  There is no such clear distinction in Upper Sco where, of the 7 millimeter detections, 4 are accreting and 3 are not.

Furthermore, as noted in Section \ref{sec:diskproperties}, the H$\alpha$ distributions of millimeter detected and non-detected sources in Upper Sco are similar.  This is a marked contrast to the younger IC348, where millimeter-detected disks are biased towards larger H$\alpha$ equivalent widths than non-detected disks.

Examining the detection statistics in greater detail, we find that the disk detection rate among CTTS in Upper Sco is $44\pm22$\%, significantly lower than that in Taurus.  The detection rate for WTTS is similar.  In Table \ref{tab:disk_compare}, we show contingency tables constructed for detections and non-detections of CTTS and WTTS in the two regions.  Using the Fisher exact test, we find for a two-tailed distribution the mm-detection rate among CTTS has $\sim0.25$\% chance of being independent of region; the apparent drop in mm-detection rate for CTTS we see in Upper Sco is significant.  For WTTS, the two-tailed Fisher's exact test indicates that the mm-detection rate for WTTS remains the same between Taurus and Upper Sco.  While the mm-detection rate for WTTS has changed little from 1 to 5 Myr, the detection rate for CTTS has decreased markedly.

\subsection{Dust mass distribution}
\label{sec:nir}

The millimeter-detection rate for Class II sources drops significantly from 1 to 5 Myr.  In Taurus, 86\% of Class II sources have millimeter detections (AW05), while in Upper Sco  that fraction has dropped to 32$\pm$13\% (6/19).  Furthermore, the median disk mass decreases.  Taurus Class II disks have a median dust mass of $3\times10^{-2}$\Mjup, while in Upper Sco  the median dust mass of detected disks is $1.3\times10^{-2}$\Mjup.  The true median is likely to be lower, due to the high censoring rate in our data.

The Class III sources in Upper Sco are unlikely to host additional massive disks.  We did not detect any of the 9 Class III disks in our survey, and the inferred upper limit to the detection rate of $<$11\% is consistent with the detection rate for Class III sources in AW05, 7$\pm$4\%.

As with the accretion state, we construct contingency tables of the detection statistics of Taurus and Upper Sco Class II and Class III sources.  We show the number of detections of each type of object in Taurus and Upper Sco in Table \ref{tab:disk_compare}.  As with CTTS, we find that the Class II sources are detected at a lower rate in Upper Sco than in Taurus.  The detection rate for Class III sources, on the other hand, is not clearly distinguishable.

\begin{figure}
  \includegraphics[width=0.95\textwidth]{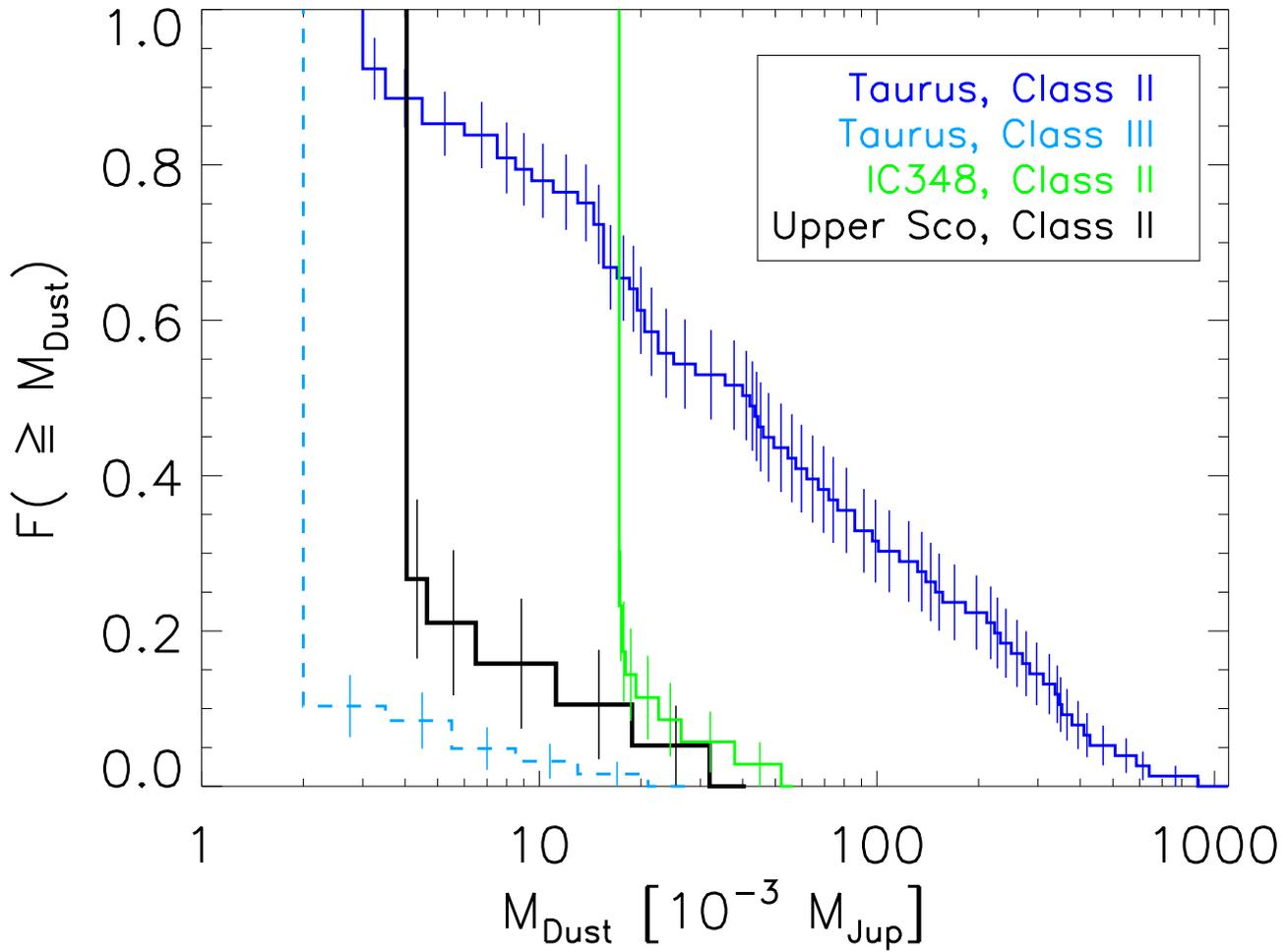}
  \caption{Cumulative dust mass distribution for Class II disks in Upper Sco, IC348, and Taurus, as well as Class III Taurus disks.  The sample distributions are calculated using the Kaplan-Meier estimator to include 3$\sigma$ upper limits.  The long vertical lines at the left indicate the lowest mass or upper limit in the sample.}
  \label{fig:mdisk}
\end{figure}

Due to the variation in exposure times in our observations, and hence sensitivities, we use the Kaplan-Meier product limit estimator for randomly censored data sets \citep{Feigelson:1985} to generate the cumulative dust mass distribution functions, F($\geq$M$_{Dust}$), for Class II sources in these regions (Fig. \ref{fig:mdisk}).  The dust mass distribution of Upper Sco Class II sources is clearly different from that of Taurus Class II sources, with the upper quartile of Upper Sco sources  overlapping the lowest quartile of Taurus sources; $27\pm10$\% of Upper Sco sources have disks with a dust mass greater than or equal to $4\times10^{-3}$\Mjup, compared to $89\pm4$\% of those in Taurus.  We make use of the Gehan, Peto-Prentice, and log-rank statistics to compare the highly censored dust mass distributions of Upper Sco and IC 348; the distributions are different, with mild statistical significance (there is only a 2.5 --7\% chance that the distributions are drawn from the same parent function).

\subsection{Disk dissipation timescale}

Having established that the dust mass distribution has changed from 1 to 5 Myr, we examine the rate at which the highest dust masses change.  Comparing similarly constructed mass distributions, LWC11 find that the fraction of IC348 disks with M$_{dust}$ greater than $2\times10^{-2}$\Mjup ~is approximately equal to the fraction of Taurus disks with M$_{dust}$ greater than $40\times10^{-2}$\Mjup, suggesting a systematic factor of 20 decrease in the dust mass of Class II sources from 1 Myr to 2.5 Myr.  If this trend were to continue, one would not expect to find any disks with millimeter continuum emission in Upper Sco.  However, our results indicate dust masses do not decline so greatly from 2.5 to 5 Myr;  the fraction of disks with M$_{dust}>2\times10^{-2}$\Mjup ~is similar between IC348 and Upper Sco, and the highest masses for Class II disks in Upper Sco are lower than masses of IC348 disks by only a factor of $\lesssim2$.

These results indicate that for Class II sources, the observed dust mass rapidly declines from 1 to 2.5 Myr, but may only slowly change from then until 5 Myr.  In contrast, the fraction of disks exhibiting IR-excess consistent with primordial dust continues its marked decline.  In Taurus, 50\% of KM stars are Class I or Class II \citep{Luhman:2010}.  We include Class I sources here as their disks consist of primordial material.  In IC348, only 39\% of KM stars are Class II, with 1\% of sources being Class I \citep[Spitzer photometry of][]{Lada:2006,Muench:2007}.  In Upper Sco the Class II fraction is 15\%, with no Class I sources.  

While there is no clear evidence for a large age spread among Upper Sco stars, \cite{Preibisch:2002} could not rule out a 1--2 Myr age spread.  This allows for the possibility of some overlap in stellar ages with the stars of IC 348 (where there may be an age spread of several Myr).  It is possible that the mm-detected sources in Upper Sco are also the youngest sources.  If these sources were only marginally older than the sources observed in IC 348, the small decrease in the maximum mm-emission for Class II disks would be consistent with a continuation of the steep drop in mm-emission seen in comparing Taurus to IC 348.

However, this argument could be equally applied to IC 348 and Taurus, with the corresponding claim that only the youngest disks are detected in the survey of LWC11, and the most massive disks in Taurus are also the youngest.  Assuming similar age spreads in these regions, this scenario would not change the general picture we find of rapid change in the greatest disk masses over several Myr, followed by a more gradual decline.

There have been recent attempts to trace the evolution of disk properties as a function of stellar age \citep[e.g.][]{2009ApJ...701..260I,Guilloteau:2011}.  The size of star forming regions, however, suggest distance uncertainties to individual stars large enough to cause 30\% or greater uncertainties in luminosity estimates; other factors (e.g. photometric errors) can make this uncertainty even higher.  In turn, 30\% luminosity uncertainties are large enough to cause age uncertainties of several Myr \citep{Hillenbrand:2008a}.  Until the age of individual stars can be accurately determined, interpretation may need to be restricted to consideration of average ages.

Comparison of HR diagrams for these regions indicates that the median age in Upper Sco is older than Taurus and IC 348 \citep[see Figure 1 of ][]{Hillenbrand:2008a}.  The comparison between Taurus and IC 348 is more ambiguous; across all temperatures, the median sequences could be equivalent within the luminosity errors.  However, at temperatures of $\log T_{eff}$ less than 3.625 (corresponding to a spectral type of about K6), the median Taurus sequence has luminosity equal to or higher than the IC 348 sequence, suggesting that the median Taurus age for KM stars is likely younger than IC 348.  These interpretations of the HR diagrams are consistent with the decrease in Class II and CTTS fraction seen from Taurus to IC 348, and then further to Upper Sco.  While the uncertainty in absolute ages of star forming regions, age spread within those regions, and ages of individual stars complicate interpretation, the overall trend is clear. 

Significant progress has been made in recent years in modeling the dust evolution in disks.  The amount of millimeter sized grains will decrease over time as dust aggregates to form centimeter and larger bodies, to which millimeter photometry is insensitive \citep{Dullemond:2005}.  \cite{Birnstiel:2010} modeled mm-slopes and fluxes in disks where millimeter-size grains remained present due to an equilibrium state between coagulation and fragmentation.  Their models are consistent with observations of young disks at 0.5--2 Myr, and later work \citep{Birnstiel:2011} shows the same mechanisms can maintain populations of micron-sized grains that produce NIR excess emission.  Observations of millimeter continuum emission from older disks, such as this work, provide important constraints on the later time evolution of dust, and perhaps constrain the mechanisms leading to loss of observable dust (e.g. radial drift, agglomeration into centimeter and larger grains).

In summary, there appears to be a rapid evolution of disks from 1 to 2.5 Myr, an interval during which both millimeter-dust masses and disk fractions rapidly drop.  While the infrared disk fraction continues to decrease from 2.5 to 5 Myr, the dust masses of the remaining disks do not rapidly change.  In addition, disks appear to reach a low baseline of dust mass where there is no longer a correlation between dust and inner disk gas.

\subsection{Implications for planet formation}
\label{sec:planetformation}

In addition to providing a point of comparison for the evolution of circumstellar disk mass as a function of age and inner disk properties, our disk dust mass measurements provide new insights on giant planet formation.

Under the assumption of a primordial 100:1 gas-to-dust mass ratio \citep[e.g.][]{Beckwith:1990} as if planet formation has yet to begin, Upper Sco retains only a single known Minimum Mass Solar Nebula disk \citep[MMSN, 10 \Mjup,][]{Weidenschilling:1977a}, the 10 \Mjup ~disk around $[$PZ99$]$ J160421.7-213028.  If planet formation were to begin now in Upper Sco, then the region could not form Jupiter mass planets at the frequency observed in the field \citep[$\sim$10\%,][]{Johnson:2009}.  However, the disks in Upper Sco are clearly evolved, and the dust has grown substantially.  Jupiter mass planets must have already formed, or be well on their way to formation, as seen in the younger sources of IC348 (LWC11).  The discovery of planetary companions to the Upper Sco members 1RXS J160929.1-210524 and GSC 06214-00210 \citep{Lafreniere:2008,Ireland:2011} serve as confirmation to this scenario.  Due to our greater sensitivity, however, we can push this argument further to examine the formation of Neptune mass planets.

At the 5 Myr age of Upper Sco, large rock and ice protoplanets may have formed, or may be in the process of formation.  There are still uncertainties in the growth of grains from centimeter to planetesimal ($10^5$ m) sizes, but simulations indicate growth from micron to centimeter size grains occurs in as little as $2\times10^5$ years \citep{Birnstiel:2010a}, and planetesimals can aggregate to form protoplanets with enough mass to accrete gas ($\sim$0.03 \Mjup) in less than 1 Myr \citep[][]{Pollack:1996}.  Under core accretion models, the formation of a gas giant planet requires that gas be retained in the disk until the accreting protoplanet reaches the critical mass ($\sim$0.1 \Mjup) at which runaway accretion begins; this process could take from less than 1 Myr \citep{Alibert:2005} to several tens of Myr \citep[][]{Pollack:1996}, depending on several factors such as disk surface density, formation radius, and the effects of planetary migration. 

Debris disks are expected to be seen in systems in the final stages of giant planet formation, as well as after the conclusion of giant planet formation.  Our results verify that BA stars in Upper Sco have no residual cold millimeter-size dust above debris levels; these disks are unlikely to retain substantial gas mass, signaling the end of giant planet formation.  

However, some lower mass KM stars have NIR or mm continuum emission and / or H$\alpha$ emission  indicating the presence of gas and dust.  The median inferred mass for the handful of mm-detected disks, under the assumption of the primordial 100:1 gas-to-dust ratio, is $\sim$0.1 MMSN.  By linear scaling of the MMSN, such a disk could produce a planet approximately twice the mass of Neptune (M$_N$ = 0.05 \Mjup) if planet formation were to begin now.  Only 4 sources have disks with inferred masses of 0.1 MMSN or greater, implying $\sim$2\% of stars in the 5 Myr Upper Sco group retain disks able to form a 2M$_N$ planet.  In comparison, results of the first 4.5 months of the Kepler mission suggest that up to 19\% of field stars have Neptune-size planet candidates within 0.5 AU \citep[][]{Borucki:2011a}.  The formation of Neptune mass planets, in addition to the formation of Jupiter mass planets, must be at an advanced stage by 5 Myr.

\section{Summary}
\label{sec:summary}

We have carried out the largest survey to date for millimeter continuum emission in a group of 5 Myr stars, an intermediate age in the evolution from ubiquitous gas and dust rich primordial disks at 1 Myr to, largely, gas and dust poor debris disks by 10 Myr.  The overall picture of Upper Sco  is of a region where a handful of primordial disks remain, with small disk masses, weak NIR excesses, and low H$\alpha$ emission.  We have found:

\begin{enumerate}
\item  5 new 1.2 millimeter detections of 2.4 mJy and brighter, with estimated dust masses typically an order of magnitude lower than Taurus disks.
\item  The sole Minimum-Mass Solar Nebula equivalent mass disk is the transition disk of $[$PZ99$]$ J160421.7-213028.
\item  The only BA star exhibiting mm-dust emission is the debris disk system with the highest 24 $\mu m$ excess, HIP 76310.
\item  Other than $[$PZ99$]$ J160421.7-213028 and HIP 76310, only Class II KM stars are detected in mm-emission.  The Class II millimeter-fraction is 32\%, a significant drop from 86\% in Taurus.    
\item  The mm-detection rate for CTTS is 44\%, significantly lower than the detection rate for CTTS in Taurus.  Millimeter detections are roughly split among accreting and non-accreting sources, suggesting the action of mechanisms that can halt accretion while preserving mm-sized grains.   
\item  Dust mass in millimeter-emitting grains appears to decrease more slowly from 2.5 to 5 Myr than from 1 to 2.5 Myr.
\item  We verify that BA stars with debris disk SEDs do not have large amounts of remnant cold dust, and find that with only four exceptions, KM stars with primordial disks lack sufficient mass to form new Neptune mass or larger planets.  If the stars of Upper Scorpius are to host giant planets at the same rate as field stars, then planet formation must be at an advanced stage.
\end{enumerate}

\begin{acknowledgments}
The authors would like to thank Jagadheep Pandian for thoughtful feedback on this paper.  G.S.M. and J.P.W. acknowledge NASA/JPL and NSF for funding support through grants RSA-1369686 and AST08-08144 respectively.  F.M., G.D., and C.P. acknowledge PNPS, CNES and ANR (contract ANR-07-BLAN-0221 and ANR-2010-JCJC-0504-01) for financial support.  C.P. acknowledges the funding from the EC 7th Framework Program as a Marie Curie Intra-European Fellow (PIEF-GA-2008-220891). \end{acknowledgments}

{\it Facilities:} \facility{IRAM:30m (MAMBO2)}

\bibliographystyle{apj}

\newpage

\newpage
\begin{deluxetable}{l l l c c c c c c c c}

  \tablecaption{Target Properties}
\tabletypesize{\scriptsize}
  \tablewidth{7.5in}
  \tablehead{
  \colhead{Name} & \colhead{RA}  &  \colhead{Dec.} & \colhead{Sp.Type\tablenotemark{a}} &   \colhead{W$_{\lambda}$(H$\alpha$)\tablenotemark{b}} &  \colhead{SED} &  \colhead{Accretion}  &  \colhead{M$_{*}$}  \\
  \colhead{}            &  \colhead{}               &  \colhead{}   &  \colhead{}                                              &  \colhead{[\AA]}                                                      &  \colhead{class} &  \colhead{state}  &  \colhead{[\Msun]} 
  }
  \startdata
HIP 76310                                                                  &      15:35:16.1 &     -25:44:03.1 &            A0V &  $<-0.1$              &  debris &  No   &   2.20 \\                                                    
HIP 77815                                                                  &      15:53:21.9 &     -21:58:16.5 &            A5V &  \nodata              &  none &    \nodata  &   2.00 \\                                                  
HIP 77911                                                                  &      15:54:41.6 &     -22:45:58.5 &            B9V &  $<-0.1$              &  debris &  No   &     2.80   \\  
HIP 78099                                                                  &      15:56:47.9 &     -23:11:02.6 &            A0V &  $<-0.1$              & none  &   No &      2.20    \\                                                     
HIP 78996                                                                  &      16:07:29.9 &     -23:57:02.3 &            A9V &  $<-0.1$              &  debris &  No  &     1.80    \\                                                     
HIP 79156                                                                  &      16:09:20.9 &     -19:27:25.9 &            A0V &  $<-0.1$              &  debris &  No  &    2.20     \\                                                     
HIP 79410                                                                  &      16:12:21.8 &     -19:34:44.6 &            B9V &  $<-0.1$              & debris  &  No  &    2.50     \\                                                    
HIP 79439                                                                  &      16:12:44.1 &     -19:30:10.2 &            B9V &  $<-0.1$              &  debris &  No  &    2.50     \\                                                    
HIP 79878                                                                  &      16:18:16.2 &     -28:02:30.1 &            A0V &  $<-0.1$              &  debris & No   &    2.30    \\                                                     
HIP 80088                                                                  &      16:20:50.2 &     -22:35:38.7 &            A9V &  $<-0.1$              & debris  &  No  &    1.70       \\                                                    
HIP 80130                                                                  &      16:21:21.1 &     -22:06:32.3 &            A9V &  $<-0.1$              &  none &   No  &      1.90    \\                                                    
     $[$PBB2002$]$ USco J155624.8-222555     &      15:56:24.8 &     -22:25:55.3 &              M4 &  -5.40, -5.51        & II           &   WTTS  &   0.24        \\                                                   
     $[$PBB2002$]$ USco J155706.4-220606     &      15:57:06.4 &     -22:06:06.1 &              M4 &  -3.60, -9.92         & II          &  WTTS   &   0.24  \\                                                   
     $[$PBB2002$]$ USco J155729.9-225843     &      15:57:29.9 &     -22:58:43.8 &              M4 &  -7.00, -6.91        & II           &  WTTS   &    0.24    \\                                                    
     $[$PBB2002$]$ USco J155829.8-231007     &      15:58:29.8 &     -23:10:07.7 &              M3 &  -250.00, -158.48 & II        &  CTTS  &      0.36     \\                                                    
     $[$PBB2002$]$ USco J160210.9-200749     &      16:02:11.0 &     -20:07:49.6 &              M5 &  -3.50                    &  III        &    WTTS    &   0.13      \\                                                     
     $[$PBB2002$]$ USco J160245.4-193037     &      16:02:45.4 &     -19:30:37.8 &              M5 &  -1.10                    &  III        &    WTTS   &   0.13     \\                                              
     $[$PBB2002$]$ USco J160357.9-194210     &      16:03:57.9 &     -19:42:10.8 &              M2 &  -3.00, -2.70         &  II        &   WTTS  &     0.49    \\                                                  
     $[$PBB2002$]$ USco J160525.5-203539     &      16:05:25.6 &     -20:35:39.7 &              M5 &  -6.10, -8.61         & III        &  WTTS   &    0.13     \\                                                     
     $[$PBB2002$]$ USco J160532.1-193315     &      16:05:32.2 &     -19:33:16.0 &              M5 &  -26.00, -152.12   & III       &   CTTS  &     0.13     \\                                                    
     $[$PBB2002$]$ USco J160545.4-202308     &      16:05:45.4 &     -20:23:08.8 &              M2 &  -35.00, -2.04       &  II        &   CTTS  &     0.49    \\                                               
     $[$PBB2002$]$ USco J160600.6-195711     &      16:06:00.6 &     -19:57:11.5 &              M5 &  -7.50, -4.11          &  II        &  WTTS   &    0.13       \\                                                
     $[$PBB2002$]$ USco J160622.8-201124     &      16:06:22.8 &     -20:11:24.4 &              M5 &  -6.00, -3.12           & II        & WTTS   &     0.13     \\                                                
     $[$PBB2002$]$ USco J160702.1-201938     &      16:07:02.1 &     -20:19:38.8 &              M5 &  -30.00, -8.30         &  II       &  CTTS   &   0.13      \\                                                     
     $[$PBB2002$]$ USco J160827.5-194904     &      16:08:27.5 &     -19:49:04.7 &              M5 &  -12.30, -14.56       & III      &  WTTS   &    0.13     \\                                                    
     $[$PBB2002$]$ USco J160900.0-190836     &      16:09:00.0 &     -19:08:36.8 &              M5 &  -15.40, -12.88       & II       &   WTTS  &    0.13       \\                           
     $[$PBB2002$]$ USco J160953.6-175446     &      16:09:53.6 &     -17:54:47.4 &              M3 &  -22.00, -22.23       & II      &   CTTS  &      0.36    \\                                                     
     $[$PBB2002$]$ USco J160959.4-180009     &      16:09:59.3 &     -18:00:09.1 &              M4 &  -4.00, -4.41            & II      &   WTTS  &    0.24      \\                                                   
     $[$PBB2002$]$ USco J161115.3-175721     &      16:11:15.3 &     -17:57:21.4 &              M1 &  -2.40, -4.47            & II       &   WTTS   &  0.60       \\                                                     
     $[$PBB2002$]$ USco J161420.3-190648     &      16:14:20.3 &     -19:06:48.1 &              K5 &  -52.00, -43.72        & II       &    CTTS  &    0.87        \\                                   
     $[$PZ99$]$ J153557.8-232405                        &      15:35:57.8 &     -23:24:04.6 &              K3 & $<$-0.1 			&    III &    WTTS      &  0.99   \\                                                
     $[$PZ99$]$ J154413.4-252258                        &      15:44:13.3 &     -25:22:59.1 &              M1 &  -3.15                        &  III   &   WTTS  &    0.60       \\                                                
     $[$PZ99$]$ J160108.0-211318                        &      16:01:08.0 &     -21:13:18.5 &              M0 &  -2.40                         & III    &   WTTS  &    0.68        \\                                                    
     $[$PZ99$]$ J160357.6-203105                        &      16:03:57.7 &     -20:31:05.5 &              K5 &  -11.57                       & II    &  CTTS   &     0.87     \\                                                    
     $[$PZ99$]$ J160421.7-213028                        &      16:04:21.7 &     -21:30:28.4 &              K2 &  -0.57          & transition     &   WTTS    &  1.12       \\                
    RX J1600.7-2343                                                    &      16:00:44.6 &     -23:43:12.0 &            M2 & \nodata                 &  III       & \nodata &    0.49     \\                                                    
    ScoPMS 31                                                            &      16:06:22.0 &     -19:28:44.6 &        M0.5V &  -21.06                  & II          &   CTTS    &  0.64      \\                                              
  \hline
$[$PBB2002$]$ USco J160823.2-193001     &      16:08:23.2 &     -19:30:00.9 &              K9 &  -6.00, -2.75            & II       &   WTTS  &   0.71    \\                                                     
$[$PBB2002$]$ USco J160900.7-190852     &      16:09:00.8 &     -19:08:52.6 &              K9 &  -12.70, -20.08        & II      &   CTTS  &    0.71       \\                                                
$[$PZ99$]$ J161411.0-230536                        &      16:14:11.1 &     -23:05:36.2 &              K0 &  0.96, 0.80, 0.38      & II     &   WTTS    &    1.35          \\            
     \hline
  \enddata
  \tablenotetext{a}{Spectral types from \cite{Hernandez:2005,1998A&A...333..619P,Preibisch:2002}}
  \tablenotetext{b}{W$_{\lambda}$(H$\alpha$) from \cite{Hernandez:2005, 1998A&A...333..619P,Preibisch:2002,2009AJ....137.4024D,Riaz:2006}}
  \label{tab:prop}
\end{deluxetable}


\begin{deluxetable}{ c | c c c | c c c | c c c  }
  \tablecaption{KM classification summary}
\tabletypesize{\scriptsize}
  \tablewidth{0pt}
  \tablehead{
       \colhead{Disk class}                &    & \colhead{CTTS}&        &  & \colhead{WTTS}  &                            &  & \colhead{Unknown H$\alpha$} &   
      }
      \startdata
          & C06 & This work    & mm & C06 & This work    & mm &  C06 & This work    & mm \\         
        \hline
        Class II   &    8    &   8     &    4                           &      11    &    11   &    2                             &     ---      &  ---       &    ---                        \\  
        Transition &    ---      &  ---       &    ---                       &      1      &    1      &    1                          &     ---      &  ---       &    ---                      \\
           Class III   &    2    &   1     &      0                         &      83    &    7   &    0                             &     22      &  1       &    0                          
      \enddata
  \tablecomments{Summary of KM star classifications, showing the number of each type in the parent sample \citep[][]{2006ApJ...651L..49C}, the number in our study, and the number detected in millimeter continuum.}
  \label{tab:disk_classification}
\end{deluxetable}


\begin{deluxetable}{l c r r}
\tabletypesize{\scriptsize}
\tablewidth{0pt} 
\tablecolumns{4} 
\tablecaption{Time on source, measured flux, and $3\sigma$ Limits}
\tablehead{
         \colhead{Name}                    &
         \colhead{ t$_{source}$}            &  
         \colhead{F$_{\nu}(1.2 mm)$}   &
         \colhead{M$_{Dust}$}\\
         \colhead{}          &                 
         \colhead{[minutes]}         &  
         \colhead{[mJy]}      &
         \colhead{[$10^{-3}$\Mjup]}
          }
\startdata
\object{HIP 76310}                                                                  & 40  & 2.41$\pm$0.61        &            3.6  \\                                                    
\object{HIP 77815}                                                                  & 30  & $<$2.00        &            $<$3.0     \\                                                  
\object{HIP 77911}                                                                  &  40  & $<$2.33      &             $<$3.5     \\                                                     
\object{HIP 78099}                                                                  & 20  & $<$2.80     &    $<$4.2   \\                                                     
\object{HIP 78996}                                                                  & 22  & $<$2.81      &   $<$4.2   \\                                                   
\object{HIP 79156}                                                                  & 32  & $<$2.18      &   $<$3.3   \\                                                   
\object{HIP 79410}                                                                 &  28  & $<$3.09     &    $<$4.6    \\                                                    
\object{HIP 79439}                                                                  &  34 & $<$2.63       &    $<$3.9  \\                                                  
\object{HIP 79878}                                                                  & 18  & $<$2.79     &    $<$4.2    \\                                                   
\object{HIP 80088}                                                                  &  28 & $<$2.47    &    $<$3.7   \\                                                    
\object{HIP 80130}                                                                 & 20  & $<$2.90     &  $<$4.4   \\                                                    
\object{$[$PBB2002$]$ J155624.8-222555}       & 10  & $<$2.47    &   $<$3.7   \\                                             
\object{$[$PBB2002$]$ J155706.4-220606}     & 22  & $<$2.88          &  $<$4.3  \\                                             
\object{$[$PBB2002$]$ J155729.9-225843}       & 20  & $<$2.46       &  $<$3.7  \\                                             
\object{$[$PBB2002$]$ J155829.8-231007}       & 20  & $<$2.29    &  $<$3.4    \\                                               
\object{$[$PBB2002$]$ J160210.9-200749}       & 20  & $<$2.44     &  $<$3.7    \\                                             
\object{$[$PBB2002$]$ J160245.4-193037}      & 24  & $<$2.41        &    $<$3.6  \\                                              
\object{$[$PBB2002$]$ J160357.9-194210}       & 20  & $<$2.45     &  $<$3.7   \\                                              
\object{$[$PBB2002$]$ J160525.5-203539}      & 10 & $<$3.58       &   $<$5.4  \\                                             
\object{$[$PBB2002$]$ J160532.1-193315}       & 20  & $<$2.58      &  $<$3.9  \\                                              
\object{$[$PBB2002$]$ J160545.4-202308}      & 30  & 5.15$\pm$0.76   &        7.7  \\                                                 
\object{$[$PBB2002$]$ J160600.6-195711}      & 10  & $<$3.24      &  $<$4.9  \\                                             
\object{$[$PBB2002$]$ J160622.8-201124}      & 20  & $<$2.88     &  $<$4.3    \\                                               
\object{$[$PBB2002$]$ J160702.1-201938}      & 20  & $<$2.45     &  $<$3.7   \\                                              
\object{$[$PBB2002$]$ J160827.5-194904}       & 10  & $<$3.53       &  $<$5.3   \\                                               
\object{$[$PBB2002$]$ J160900.0-190836}     & 10  & $<$3.19      &  $<$4.8 \\                                               
\object{$[$PBB2002$]$ J160953.6-175446}     & 10  & $<$3.00      & $<$4.5    \\                                               
\object{$[$PBB2002$]$ J160959.4-180009}    &  10 & $<$3.43       &$<$5.1  \\                                               
\object{$[$PBB2002$]$ J161115.3-175721}       & 10  & $<$4.23       & $<$6.3  \\                                               
\object{$[$PBB2002$]$ J161420.3-190648}            & 10  & 9.79$\pm$1.14     &     14.7  \\                     
\object{$[$PZ99$]$ J153557.8-232405}                   & 20  & $<$2.78     &      $<$4.2  \\                               
\object{$[$PZ99$]$ J154413.4-252258}                        & 20  & $<$2.78      &  $<$4.2  \\                                         
\object{$[$PZ99$]$ J160108.0-211318}                          & 20  & $<$2.68      &$<$4.0  \\                                            
\object{$[$PZ99$]$ J160357.6-203105}                        & 20  & $<$2.47      & $<$3.7    \\                                               
\object{$[$PZ99$]$ J160421.7-213028}                        & 8  & 67.51$\pm$1.44    &      101.3  \\                                            
\object{RX J1600.7-2343}                                                      & 20  & $<$2.50       &   $<$3.8  \\                                             
\object{ScoPMS 31}                                                               & 50  & 2.70$\pm$0.61    &           4.1  \\                     
  \hline
$[$PBB2002$]$ USco J160823.2-193001     &      ---		&  21.90$\pm$4.70\tablenotemark{a}       &   40.9  \\                                                     
$[$PBB2002$]$ USco J160900.7-190852     &       ---	&    12.30$\pm$2.00\tablenotemark{a}       &   23.0 \\                                                
$[$PZ99$]$ J161411.0-230536                        &       ---	&   3.5$\pm$1.1\tablenotemark{b}     &  6.0   \\            
                           
\enddata
  \tablenotetext{a}{1.3 $mm$ flux from \cite{Cieza:2008}.}
  \tablenotetext{b}{1.2 $mm$ flux from \cite{2009A&A...497..409R}.}
\label{tab:obs}
\end{deluxetable}


\begin{deluxetable}{ c | c c | c c || c c | c c   }
  \tablecaption{Comparisons with Taurus disks}
\tabletypesize{\scriptsize}
  \tablewidth{0pt}
  \tablehead{
       \colhead{~} &   \colhead{CTTS}  &    & \colhead{WTTS}  &   &    \colhead{Class II} &   & \colhead{Class III}   
      }
      \startdata
        	mm-detected?	  	& yes 	& no    	&  yes 	& no 		&	yes	&   no	&  yes	&  no	\\         
        \hline
          Taurus   			&  67		&  	7	&	9	&	52	&	64	&  10		&  4		& 	50	\\
          Upper Sco   		&  4		&	5	&	3	&	16	&	6	&  13	 	&  0		&	9	\\		     
          \hline
          P($f_{Tau} = f_{US}$)	&   0.0025	&	&  \textbf{$\approx1$}	&		& $<<$0.001 &		&  \textbf{$\approx1$}	&		\\   
      \enddata
  \tablecomments{Contingency tables for comparing detection statistics of CTTS / WTTS and Class II / Class III sources in Upper Sco and Taurus.  P($f_{Tau} = f_{US}$) shows the result of the two-tailed Fisher exact test for each table, showing the probability for the mm-detection rate in Taurus being equal to the mm-detection rate in Upper Sco for each type of object.}
  \label{tab:disk_compare}
\end{deluxetable}

\end{document}